%% file: conference_101719.tex
\documentclass[conference]{IEEEtran}
\IEEEoverridecommandlockouts
\usepackage{cite}
\usepackage{amsmath,amssymb,amsfonts}
\usepackage{algorithmic}
\usepackage{graphicx}
\usepackage{textcomp}
\usepackage{xcolor}
\def\BibTeX{{\rm B\kern-.05em{\sc i\kern-.025em b}\kern-.08em
    T\kern-.1667em\lower.7ex\hbox{E}\kern-.125emX}}

\usepackage{amssymb}
\usepackage{lipsum}
\usepackage{amsthm}
\usepackage{amsmath}
\usepackage{gensymb}
\usepackage{float}
\usepackage{graphicx}
\usepackage{subcaption}
\usepackage{tabularx}
\usepackage{comment}
\usepackage{dblfloatfix}
\usepackage[table,xcdraw]{xcolor}
\usepackage{wrapfig}
\usepackage{soul}
\usepackage{lipsum} 
\usepackage{tikz}
\tikzset{
  dashdot/.style={dash pattern=on 3pt off 2pt on 0.8pt off 2pt},
  dashdotthick/.style={dash pattern=on 3pt off 2pt on 0.8pt off 2pt, line cap=round}
}

\usepackage{xcolor}

\begin{document}

\title{Integrating Active Damping with Shaping-Filtered Reset Tracking Control for Piezo-Actuated Nanopositioning}

\author{\IEEEauthorblockN{Aditya Natu}
\IEEEauthorblockA{\textit{Precision and Microsystems Engineering} \\
\textit{Delft University of Technology}\\
Delft, Netherlands \\
A.M.Natu@tudelft.nl}
\and
\IEEEauthorblockN{Xiaozhe Hu}
\IEEEauthorblockA{\textit{Precision and Microsystems Engineering} \\
\textit{Delft University of Technology}\\
Delft, Netherlands \\
X.Hu-17@student.tudelft.nl}
\and
\IEEEauthorblockN{Hassan HosseinNia}
\IEEEauthorblockA{\textit{Precision and Microsystems Engineering} \\
\textit{Delft University of Technology}\\
Delft, Netherlands \\
S.H.HosseinNiaKani@tudelft.nl}

}

\maketitle

\begin{abstract}
Piezoelectric nanopositioning systems are often limited by lightly damped structural resonances and the gain--phase constraints of linear feedback, which restrict achievable bandwidth and tracking performance. This paper presents a dual-loop architecture that combines an inner-loop non-minimum-phase resonant controller (NRC) for active damping with an outer-loop tracking controller augmented by a constant-gain, lead-in-phase (CgLp) reset element to provide phase lead at the targeted crossover without increasing loop gain. We show that aggressively tuned CgLp designs with larger phase lead can introduce pronounced higher-order harmonics, degrading error sensitivity in specific frequency bands and causing multiple-reset behavior. To address this, a shaping filter is introduced in the reset-trigger path to regulate the reset action and suppress harmonic-induced effects while preserving the desired crossover-phase recovery. The proposed controllers are implemented in real time on an industrial piezo nanopositioner, demonstrating an experimental open-loop crossover increase of approximately 55~Hz and a closed-loop bandwidth improvement of about 34~Hz relative to a well-tuned linear baseline.
\end{abstract}


\definecolor{matlabgray}{rgb}{0.4,0.4,0.4}        
\definecolor{matlabyellow}{rgb}{0.9290,0.6940,0.1250}
\definecolor{matlaborange}{rgb}{0.8500,0.3250,0.0980}
\definecolor{matlabblue}{rgb}{0,0.4470,0.7410}
\definecolor{matlabgreen}{rgb}{0.4660,0.6740,0.1880}
\definecolor{matlabmaroon}{rgb}{0.6350,0.0780,0.1840}
\definecolor{matlabpurple}{rgb}{0.4940,0.1840,0.5560}
\definecolor{matlablightgray}{rgb}{0.8,0.8,0.8}   

\newcommand{\dashdotstyle}{dash pattern=on 3pt off 2pt on 1pt off 2pt}

\newcommand{\blackline}{%
  \raisebox{2pt}{\tikz{\draw[-,black,line width=1.5pt](0,0)--(3mm,0);}}}
\newcommand{\blacklinedashed}{%
  \raisebox{2pt}{\tikz{\draw[-,black,dashed,line width=1pt](0,0)--(3mm,0);}}}
\newcommand{\blacklinedashdot}{%
  \raisebox{2pt}{\tikz{\draw[-,black,\dashdotstyle,line width=1pt](0,0)--(3mm,0);}}}
\newcommand{\blacklinedotted}{%
  \raisebox{2pt}{\tikz{\draw[-,black,dotted,line width=1pt](0,0)--(3mm,0);}}}

\newcommand{\grayline}{%
  \raisebox{2pt}{\tikz{\draw[-,matlabgray,line width=1.5pt](0,0)--(3mm,0);}}}
\newcommand{\graylinedashed}{%
  \raisebox{2pt}{\tikz{\draw[-,matlabgray,dashed,line width=1pt](0,0)--(3mm,0);}}}
\newcommand{\graylinedashdot}{%
  \raisebox{2pt}{\tikz{\draw[-,matlabgray,\dashdotstyle,line width=1pt](0,0)--(3mm,0);}}}
\newcommand{\graylinedotted}{%
  \raisebox{2pt}{\tikz{\draw[-,matlabgray,dotted,line width=1pt](0,0)--(3mm,0);}}}

\newcommand{\lightgrayline}{%
  \raisebox{2pt}{\tikz{\draw[-,matlablightgray,line width=1.5pt](0,0)--(3mm,0);}}}
\newcommand{\lightgraylinedashed}{%
  \raisebox{2pt}{\tikz{\draw[-,matlablightgray,dashed,line width=1pt](0,0)--(3mm,0);}}}
\newcommand{\lightgraylinedashdot}{%
  \raisebox{2pt}{\tikz{\draw[-,matlablightgray,\dashdotstyle,line width=1pt](0,0)--(3mm,0);}}}
\newcommand{\lightgraylinedotted}{%
  \raisebox{2pt}{\tikz{\draw[-,matlablightgray,dotted,line width=1pt](0,0)--(3mm,0);}}}

\newcommand{\yellowline}{%
  \raisebox{2pt}{\tikz{\draw[-,matlabyellow,line width=1.5pt](0,0)--(3mm,0);}}}
\newcommand{\yellowlinedashed}{%
  \raisebox{2pt}{\tikz{\draw[-,matlabyellow,dashed,line width=1pt](0,0)--(3mm,0);}}}
\newcommand{\yellowlinedashdot}{%
  \raisebox{2pt}{\tikz{\draw[-,matlabyellow,\dashdotstyle,line width=1pt](0,0)--(3mm,0);}}}
\newcommand{\yellowlinedotted}{%
  \raisebox{2pt}{\tikz{\draw[-,matlabyellow,dotted,line width=1pt](0,0)--(3mm,0);}}}

\newcommand{\orangeline}{%
  \raisebox{2pt}{\tikz{\draw[-,matlaborange,line width=1.5pt](0,0)--(3mm,0);}}}
\newcommand{\orangelinedashed}{%
  \raisebox{2pt}{\tikz{\draw[-,matlaborange,dashed,line width=1pt](0,0)--(3mm,0);}}}
\newcommand{\orangelinedashdot}{%
  \raisebox{2pt}{\tikz{\draw[-,matlaborange,\dashdotstyle,line width=1pt](0,0)--(3mm,0);}}}
\newcommand{\orangelinedotted}{%
  \raisebox{2pt}{\tikz{\draw[-,matlaborange,dotted,line width=1pt](0,0)--(3mm,0);}}}

\newcommand{\blueline}{%
  \raisebox{2pt}{\tikz{\draw[-,matlabblue,line width=1.5pt](0,0)--(3mm,0);}}}
\newcommand{\bluelinedashed}{%
  \raisebox{2pt}{\tikz{\draw[-,matlabblue,dashed,line width=1pt](0,0)--(3mm,0);}}}
\newcommand{\bluelinedashdot}{%
  \raisebox{2pt}{\tikz{\draw[-,matlabblue,\dashdotstyle,line width=1pt](0,0)--(3mm,0);}}}
\newcommand{\bluelinedotted}{%
  \raisebox{2pt}{\tikz{\draw[-,matlabblue,dotted,line width=1pt](0,0)--(3mm,0);}}}

\newcommand{\greenline}{%
  \raisebox{2pt}{\tikz{\draw[-,matlabgreen,line width=1.5pt](0,0)--(3mm,0);}}}
\newcommand{\greenlinedashed}{%
  \raisebox{2pt}{\tikz{\draw[-,matlabgreen,dashed,line width=1pt](0,0)--(3mm,0);}}}
\newcommand{\greenlinedashdot}{%
  \raisebox{2pt}{\tikz{\draw[matlabgreen,dashdot,line width=1pt](0,0)--(4mm,0);}}}
\newcommand{\greenlinedotted}{%
  \raisebox{2pt}{\tikz{\draw[-,matlabgreen,dotted,line width=1pt](0,0)--(3mm,0);}}}

\newcommand{\maroonline}{%
  \raisebox{2pt}{\tikz{\draw[-,matlabmaroon,line width=1.5pt](0,0)--(3mm,0);}}}
\newcommand{\maroonlinedashed}{%
  \raisebox{2pt}{\tikz{\draw[-,matlabmaroon,dashed,line width=1pt](0,0)--(3mm,0);}}}
\newcommand{\maroonlinedashdot}{%
  \raisebox{2pt}{\tikz{\draw[-,matlabmaroon,\dashdotstyle,line width=1pt](0,0)--(3mm,0);}}}
\newcommand{\maroonlinedotted}{%
  \raisebox{2pt}{\tikz{\draw[-,matlabmaroon,dotted,line width=1pt](0,0)--(3mm,0);}}}

\newcommand{\purpleline}{%
  \raisebox{2pt}{\tikz{\draw[-,matlabpurple,line width=1.5pt](0,0)--(3mm,0);}}}
\newcommand{\purplelinedashed}{%
  \raisebox{2pt}{\tikz{\draw[-,matlabpurple,dashed,line width=1pt](0,0)--(3mm,0);}}}
\newcommand{\purplelinedashdot}{%
  \raisebox{2pt}{\tikz{\draw[matlabpurple,dashdot,line width=1pt](0,0)--(4mm,0);}}}
\newcommand{\purplelinedotted}{%
  \raisebox{2pt}{\tikz{\draw[-,matlabpurple,dotted,line width=1pt](0,0)--(3mm,0);}}}

\input{Text/Introduction}

\input{Text/Preliminaries}
\input{Text/SystemDescription}
\input{Text/Methodology}

\input{Text/Simulations}

\input{Text/Experimental}
\input{Text/Conclusions}
\input{Text/Statements}

\bibliographystyle{IEEEtran}
\bibliography{References}

\end{document}

%% file: Text/Introduction.tex
\section{Introduction}
\label{sec:Introduction}

Piezoelectric nanopositioning systems are widely used in applications including AFM probe positioning \cite{fleming2014design}, semiconductor alignment \cite{yao2017line}, and biomedical manipulation \cite{xu2016advanced}, where both nanometer-scale precision and high-speed motion are essential. These systems typically employ piezoelectric stack actuators, which provide high stiffness and large force capability, enabling high resolution and potentially wide bandwidth. The motion stage is commonly guided by parallel flexures, ensuring backlash-free and frictionless operation \cite{ru2016nanopositioning}. However, the resulting compliant structure often exhibits lightly damped resonant modes that fundamentally limit the achievable closed-loop bandwidth and disturbance rejection \cite{devasia2007survey}.

In practice, proportional–integral (PI) controllers remain a common baseline for feedback control. However, the dominant resonance associated with lightly damped flexures typically restricts the control bandwidth to a fraction of the resonance frequency \cite{fleming2009nanopositioning}. To address these limitations, a range of active damping strategies have been proposed to suppress lightly damped structural modes, including Positive Position Feedback (PPF) \cite{li2015positive} and Integral Resonant Control (IRC) \cite{al2013integral}. These approaches introduce targeted damping around resonance frequencies and are often deployed in dual-loop architectures, where an inner damping loop increases effective structural damping and an outer motion-control loop achieves high tracking bandwidth \cite{chen2021damping}. More recently, the non-minimum-phase resonant controller (NRC) \cite{NATU2026106790} has been introduced, enabling substantially higher closed-loop bandwidths, including operation beyond the first dominant resonance.

Despite these bandwidth improvements, linear control remains fundamentally limited by the waterbed effect and the Bode gain–phase relationship. As the crossover frequency approaches the first structural resonance, the associated phase lag and gain peaking degrade tracking performance and can reduce closed-loop stability and robustness \cite{sebastian2005design}. Classical phase-lead compensation can partially alleviate the phase deficit, but typically at the cost of increased high-frequency gain, which is undesirable due to noise amplification.

Nonlinear reset control has shown a strong potential to overcome these limitations by partially breaking the waterbed effect and extending the achievable performance of precision motion systems \cite{saikumar2021loop}. A key advantage of reset control is the availability of frequency-domain analysis tools for open- and closed-loop design. Canonical reset elements such as the Clegg integrator and the first-order reset element (FORE) provide reduced phase lag while largely preserving the gain characteristics of their linear counterparts \cite{clegg1958nonlinear,guo2009frequency}. Building on these concepts, the constant-gain, lead-in-phase (CgLp) element combines a generalized FORE (GFORE) with a linear lead–lag filter to deliver approximately constant gain with phase lead. CgLp has been successfully applied to improve robust phase margins and enable higher low-frequency loop gains in industrial motion applications, including wire bonding and precision positioning systems \cite{saikumar2019constant,hosseini2025frequency}. However, a more aggressive reset action increases nonlinearity, leading to stronger higher-order harmonics that can degrade system performance. These harmonics can cause multiple zero-crossings in the reset signal, invalidating frequency-domain analysis tools such as higher-order sinusoidal-input describing functions (HOSIDFs), which assume only two resets per time period \cite{zhang2024frequency}.

To reduce reset-induced harmonics, prior work has investigated continuous-reset CgLp (CRCgLp) and shaping-filter approaches that attenuate higher-order harmonics within targeted frequency bands \cite{karbasizadeh2023shaping}. CRCgLp utilizes pre- and post-filtering with linear elements to reduce higher-order harmonics and improve transient performance \cite{karbasizadeh2022continuous}. However, CRCgLp relies on a lead element that can amplify measurement noise in the error signal driving the reset action, thereby increasing HOSIDFs \cite{cai2020optimal}. The shaping filter is a technique that independently shapes the signal that triggers the reset, reducing higher-order harmonics in the reset signal without affecting the control signal, thereby eliminating multiple resets \cite{zhang2025enhancing}.

Combining linear active damping (to suppress structural resonances) with nonlinear reset control (to enhance tracking) is therefore a promising strategy. Recent results have explored CgLp-based reset control together with PPF damping, including CRCgLp to further limit nonlinearities and improve tracking \cite{hakvoort2023frequency}.

In this study, we investigate a dual-loop architecture that integrates an NRC-based active damping controller with CgLp-based reset control to increase open- and closed-loop bandwidth and improve tracking performance in a piezoelectric nanopositioning system. The reset element is tuned across multiple phase-lead configurations to quantify the achievable performance gains. For higher phase-lead settings, reset-induced higher-order harmonics become pronounced over specific frequency ranges, causing performance degradation and multiple-reset events. To address this, a shaping filter is introduced to suppress the problematic harmonics and prevent excessive resetting. The proposed controllers are implemented in real time on an industrial nanopositioner, and performance is evaluated experimentally in both the frequency domain and the time domain, demonstrating improved open- and closed-loop characteristics relative to a baseline linear design.

The remainder of the paper is organized as follows. Section~\ref{sec:Preliminaries} introduces the required preliminaries, Section~\ref{sec:ControlDesign} presents the control architecture and design requirements, Section~\ref{sec:Simulations} reports the simulation study and shaping-filter design, Section~\ref{sec:Experimental} presents the experimental frequency responses and tracking results, and Section~\ref{sec:Conclusion} concludes the paper.

%% file: Text/Preliminaries.tex
\section{Preliminaries}
\label{sec:Preliminaries}
\subsection{Reset Control}
The reset element, denoted by $\mathcal{R}$, is a time-invariant hybrid system. Its state-space representation, with an input signal $e_r(t)$, an output signal $u_r(t)$, and a state vector $x_r(t) \in \mathbb{R}^{n_r \times 1}$, is defined as follows
\begin{equation}
\label{Eq:ResetElement}
\mathcal{R}: \begin{cases}\dot{x}_r(t)=A_r x_r(t)+B_r e_r(t), & \left(x_r(t), e_r(t)\right) \notin \mathcal{F}, \\ x_r\left(t^{+}\right)=A_\rho x_r(t), & \left(x_r(t), e_r(t)\right) \in \mathcal{F}, \\ u_r(t)=C_r x_r(t)+D_r e_r(t), & \end{cases}
\end{equation}
where the reset surface $\mathcal{F}$ is given by:
\begin{equation}
\mathcal{F}:=\left\{e_r(t)=0 \wedge\left(A_\rho-I\right) x_r(t) \neq 0\right\},
\end{equation}
with $A_r \in \mathbb{R}^{n_r \times n_r}, B_r \in \mathbb{R}^{n_r \times 1}, C_r \in \mathbb{R}^{1 \times n_r}$, and $D_r \in \mathbb{R}$ the state-space matrices of the reset element, and the reset value matrix is denoted by $A_\rho= \operatorname{diag}\left(\gamma_1, \ldots, \gamma_{n_r}\right)$. $x_r\left(t^{+}\right)\in \mathbb{R}^{n_r \times 1}$ is the after-reset state vector. By defining the base linear system of the reset element as the case when there is no reset, its transfer function is calculated as follows:
\begin{equation}
    \mathcal{R}_{bl}(s)=C_r\left(s-A_r\right)^{-1} B_r+D_r,
\end{equation}
where $s \in \mathbb{C}$ is the Laplace variable.

\subsection{First-Order Reset Element}
In this study, a generalized first-order reset element (GFORE) is utilized,  with its matrices expressed as:
\begin{equation}
\begin{aligned}
& A_r=-\omega_\alpha, B_r=\omega_\beta, C_r=1, D_r=0 \\
& A_\rho=\gamma_r \in(-1,1), \text { where } \omega_\alpha \geq 0 \in \mathbb{R}, \omega_\beta \in \mathbb{R}^{+} .
\end{aligned}
\end{equation}

\subsection{Higher-Order Sinusoidal-Input Describing Functions}
The frequency domain analysis tool, known as Higher-Order Sinusoidal-Input Describing Functions (HOSIDFs) is introduced for reset controllers to enable a deeper analysis considering higher order harmonics \cite{saikumar2021loop}. Considering $e_r(t)=\sin (\omega t)$ as the input of the reset element, then for its output, we have
\begin{equation}
    u_r(t)=\sum_{n=1}^{\infty}\left|H_n(\omega)\right| \sin \left(n \omega t+\angle H_n(\omega)\right),
\end{equation}
where $H_n(\omega)$ is the HOSIDF of the reset element in Eq. (\ref{Eq:ResetElement}) as follows:
\begin{equation}
\resizebox{\columnwidth}{!}{$
H_n(\omega)=
\begin{cases}
C_r\left(j \omega I-A_r\right)^{-1}\left(I+j \Theta_D(\omega)\right) B_r+D_r, & \text { for } n=1, \\
C_r\left(j n \omega I-A_r\right)^{-1} j \Theta_D(\omega) B_r, & \text { for odd } n \geq 2, \\
0, & \text { for even } n \geq 2,
\end{cases}
$}
\end{equation}

with
\begin{equation}
\begin{aligned}
& \Lambda(\omega)=\omega^2 I+A_r^2, \quad \Delta(\omega)=I+e^{\frac{\pi}{\omega} A_r} \\
& \Delta_r(\omega)=I+A_\rho e^{\frac{\pi}{\omega} A_r}, \quad \Gamma_r(\omega)=\Delta_r^{-1}(\omega) A_\rho \Delta(\omega) \Lambda^{-1}(\omega) \\
& \Theta_{\mathrm{D}}(\omega)=-\frac{2 \omega^2}{\pi} \Delta(\omega)\left[\Gamma_r(\omega)-\Lambda^{-1}(\omega)\right] .
\end{aligned}
\end{equation}

%% file: Text/SystemDescription.tex
\section{System Description and Identification}
\label{sec:SystemDescription}

\begin{figure}[t!]
    \centering
    \subfloat[]{\includegraphics[width=0.7\columnwidth]{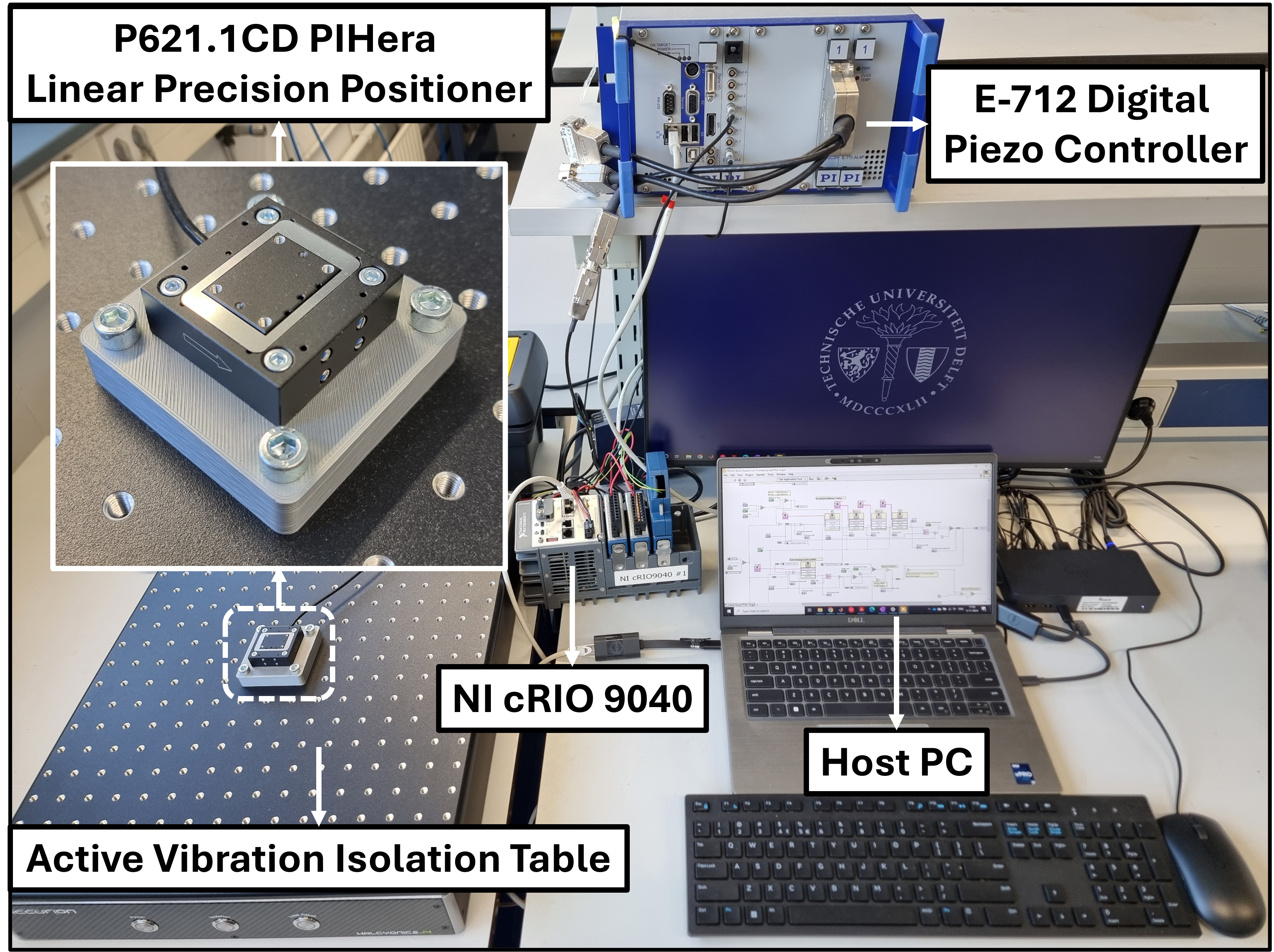}
    \label{fig:ExpPlatform}}
    \hfil
    \subfloat[]{\includegraphics[width=0.85\columnwidth]{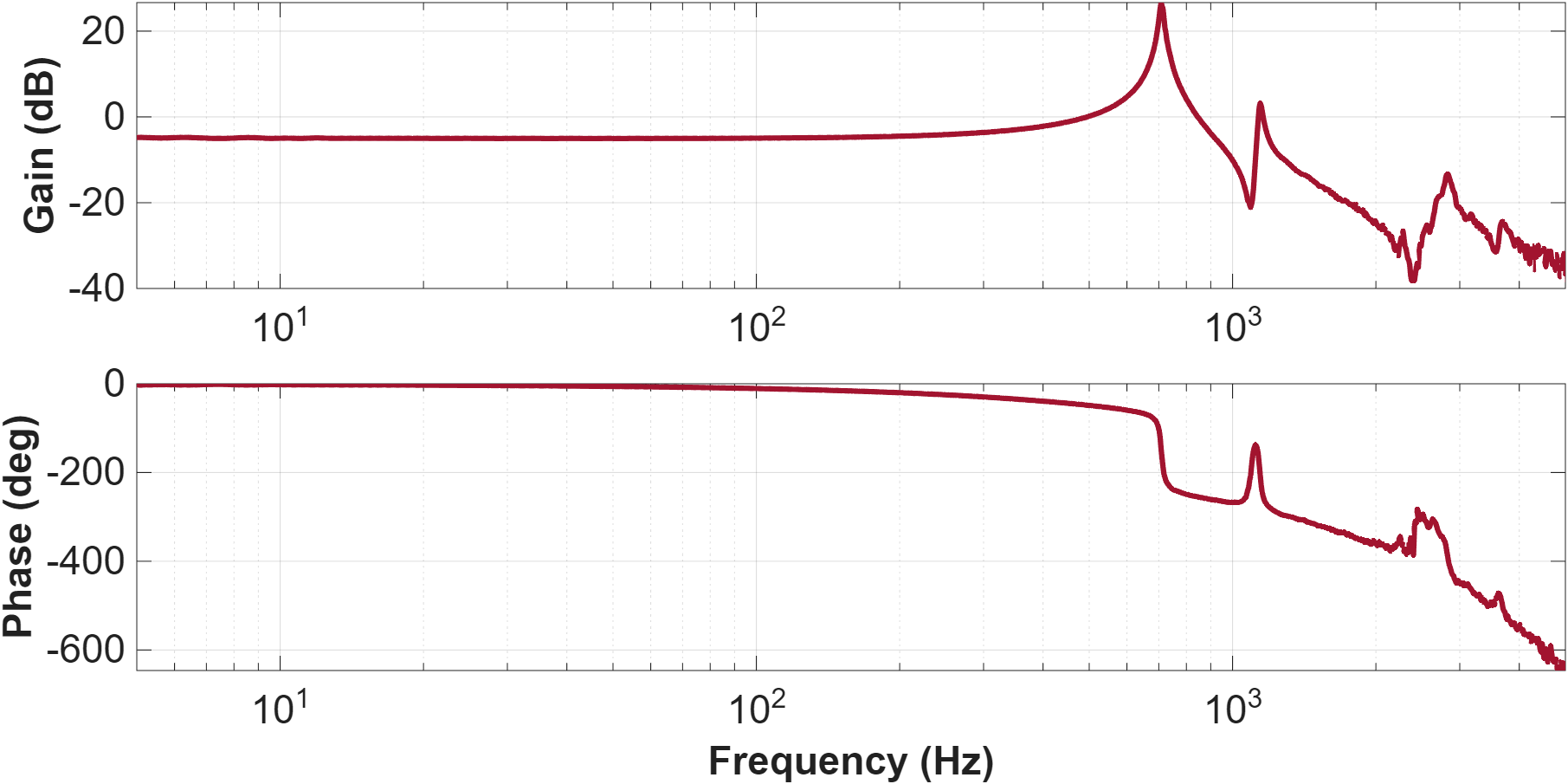}
    \label{fig:SystemIdentification}}
    \caption{(a) Experimental setup of a piezo-actuated nanopositioning system, (b) Identified frequency response of the nanopositioning system $G(s)$.}
    \label{fig:ExpSetup_SysIden}
\end{figure}

As shown in Fig. \ref{fig:ExpSetup_SysIden}(a), the experimental setup uses a commercial P-621.1CD PIHera single-axis nanopositioner (travel range: 100~$\mu$m). The stage includes a ceramic-insulated multilayer piezo-stack actuator, a flexure-guided mechanism, and a high-resolution capacitive sensor. Actuation and sensing are managed by an E-712 piezo-controller with integrated voltage amplification and sensor conditioning. The hardware is interfaced with an NI CompactRIO chassis with an embedded FPGA for external control, and 16-bit analog I/O modules for signal transmission and acquisition. The control scheme is implemented in NI LabVIEW, which links the host computer and the nanopositioner. The actuation voltage range is 0–10~V, and the FPGA sampling time is $t_s=30~\mu$s ($f_s=33.\overline{33}$~kHz), giving a Nyquist frequency of $\approx16.66$~kHz, more than six times the 0–3000~Hz frequency range of interest for system identification and control. Controller discretization uses the bilinear (Tustin) transform. The FPGA uses 64-bit double-precision floating-point arithmetic ($\approx$15 decimal digits), and the LabVIEW code is optimized so the FPGA loop rate matches the sampling rate, preventing data loss.

A sinusoidal chirp (0.75–1 V) was generated in LabVIEW and applied to the piezo-actuator for system identification. The capacitive sensor measured the position, and the input-output data were imported into MATLAB for analysis. Transfer functions were estimated using MATLAB's signal processing toolbox. A sampling frequency \(F_s = 33.3\) kHz provided sufficient data for accurate identification.
The dominant resonance \( \omega_n \) is at 710 Hz, with a nearby second mode \( \omega_2 \) at 1150 Hz (see Fig. \ref{fig:ExpSetup_SysIden}(b)). A significant phase delay, due to actuator-amplifier dynamics and system time delay, appears even below \( \omega_n \). Higher modes occur at and above 2000 Hz. Pole-zero interlacing confirms that the system is collocated.

%% file: Text/Methodology.tex
\section{Control Methodology}
\label{sec:ControlDesign}

\subsection{Control Architecture}

\begin{figure}[t!]
    \centering
    \includegraphics[width=1\linewidth]{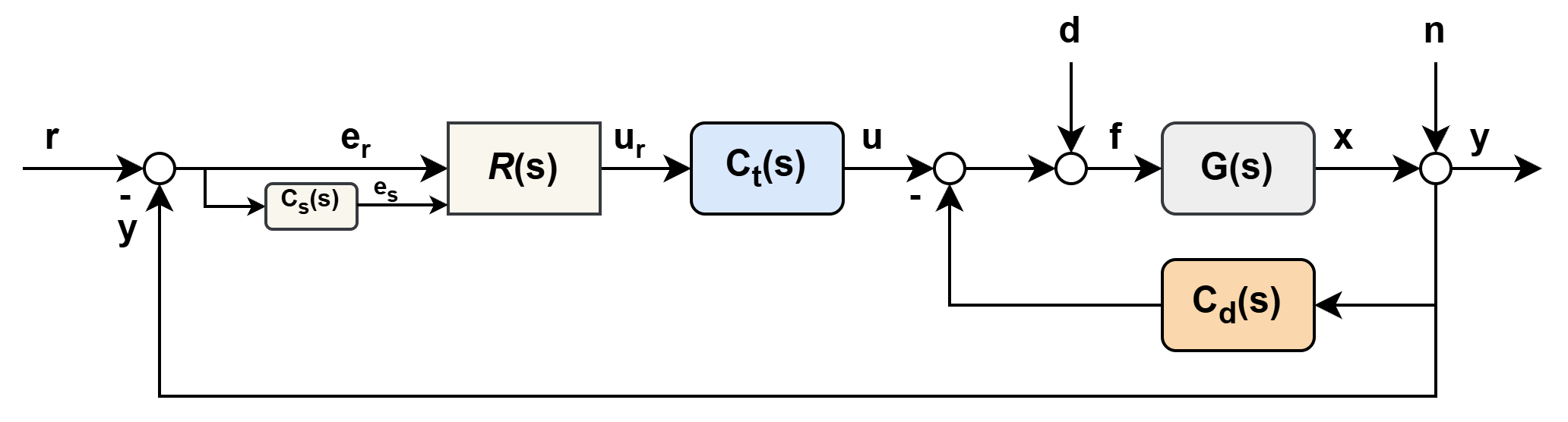}
    \caption{Dual closed-loop architecture with damping controller $C_d(s)$ and tracking controller $C_t(s)$ incorporating a reset element $\mathcal{R}(s)$.}
    \label{fig:ControlArchitecture}
\end{figure}

The dual-loop control architecture in Fig.~\ref{fig:ControlArchitecture} comprises an inner active damping loop with a linear controller $C_d(s)$ and an outer tracking loop with a linear controller $C_t(s)$. The inner loop attenuates the dominant resonant modes of the plant, enabling the outer loop to operate at a higher closed-loop bandwidth and improve tracking performance.

The system is driven by the reference input \(r\), while the process disturbance \(d\) and the output disturbance \(n\) act on the dynamics of the system. The measurement noise is included within \(n\), thus the output \(y\) corresponds to the measured position. The actual stage position is denoted by \(x\). 

The outer control loop incorporates a nonlinear reset element \(\mathcal{R}\) connected in series with \(C_t(s)\). This reset element is driven by the error signal \(e_r\) and generates the reset control signal \(u_r\). Furthermore, a shaping filter \(C_s(s)\) is implemented in the reset path to produce the reset-triggering signal \(e_s\) and to modulate the reset action.

\subsection{Control Design Requirements}
For the control architecture in Fig. \ref{fig:ControlArchitecture}, the dual open-loop transfer function is given as: 
\begin{equation}
\begin{aligned}
L_{D_1}(j \omega) & =G(j \omega)\left(H_1(j \omega) C_t(j \omega)+C_d(j \omega)\right) \\
L_{D_{n>1}}(j \omega) & =G(n j \omega)\left(H_n(j \omega) C_t(j n \omega)+C_d(j n \omega)\right)
\end{aligned}
\end{equation}
while the outer open-loop transfer function is given as: 
\begin{equation}
\label{Eq:L1}
\begin{aligned}
L_1(j \omega) & =H_1(j \omega) C_t(j \omega) \frac{G(j \omega)}{1+G(j \omega) C_d(j \omega)} \\
L_{n>1}(j \omega) & =H_n(j \omega) C_t(j n \omega) \frac{G(j n \omega)}{1+G(j n \omega) C_d(j n \omega)}
\end{aligned}
\end{equation}

The HOSIDF-based sensitivity mapping from the reference to the error is defined as:
\begin{equation}
\resizebox{\columnwidth}{!}{$
\begin{aligned}
S_{er_1}(j \omega) & =\frac{E_1}{R}=\frac{1}{1+L_1(j \omega)} \\
S_{er_{n>1}}(j \omega) & =\frac{E_n}{R}=-L_n(j \omega)\, S_{bl}(j n \omega)
\left(\left|S_{er_1}(j \omega)\right| \angle\left(n \angle S_{er_1}(j \omega)\right)\right)
\end{aligned}
$}
\end{equation}

where
\begin{equation}
\begin{aligned}
S_{bl}(j n \omega) & =\frac{1}{1+L_{b l}(j n \omega)} \\
L_{b l}(j n \omega) & =\mathcal{R}_{b l}(j n \omega) \frac{G(j n \omega)}{1+G(j n \omega) C_d(j n \omega)}
\end{aligned}
\end{equation}

Similarly, the HOSIDF-based closed-loop transfer function mapping reference to output is expressed as:
\begin{equation}
\resizebox{\columnwidth}{!}{$
\begin{aligned}
T_{yr_1}(j\omega) &= \frac{Y_1}{R}=L_1(j\omega)\,S_{er_1}(j\omega) \\
T_{yr_{n>1}}(j\omega) &= \frac{Y_n}{R}
= L_n(j\omega)\,S_{bl}(jn\omega)\left(\left|S_{er_1}(j\omega)\right|\angle\left(n\angle S_{er_1}(j\omega)\right)\right)
\end{aligned}
$}
\end{equation}

The open-loop bandwidth frequency $\omega_b \in \mathbb{R}^{+}$ of a reset control system is defined as the frequency at which the magnitude of the first-order harmonic open-loop transfer function ${L}_1(j\omega)$, as given in Eq. (\ref{Eq:L1}), reaches 0 dB, mathematically expressed as:
\begin{equation}
|{L}_1\left(j\omega_b\right)|=0 \mathrm{~dB} .
\end{equation}

In this work, a well-designed linear controller case (when $\mathcal{R}_{bl}(s) = 1$) will be compared to different cases where a reset element is employed in the outer loop in conjunction with the tracking controller $C_t(s)$ to improve the system performance. The performance objective is maximizing open-loop control bandwidth $\omega_b$, mathematically expressed as follows: 

\begin{equation}
    \label{Eq_Objective1}
        \max \hspace{1mm} \omega_b \hspace{3mm} \forall \hspace{1mm} \omega_b\in \mathbb{R}^{+}.
    \end{equation}

Subsequently, it will be shown that achieving this performance objective improves the closed-loop behavior, as reflected by the sensitivity functions $S_{er}(s)$ and $T_{yr}(s)$.

\subsection{Controller Design}

To suppress the first dominant resonance, in this work, a non-minimum-phase resonant controller (NRC) is employed for active damping, defined as
\begin{equation}
\label{Eq:NRC_Controller}
C_{d}(s)
= k \left(\frac{s - \omega_{a}}{s + \omega_{a}}\right),
\end{equation}
where $k$, and $\omega_{a}$ denote the controller gain and corner frequency, respectively. The controller parameters are selected as $k = \gamma \cdot |G(0)|^{-1}$ and $\omega_{a} = n\cdot\omega_{n}$, where $\omega_{n}$ is the first resonant frequency of the system. Due to its non-minimum-phase structure, the NRC provides gain–phase decoupling, enabling the effective suppression of the targeted resonance mode  \cite{NATU2026106790}.

For the outer loop, the tracking controller is designed as a series combination of a PI controller, notch filters targeting the dominant higher-order modes, and a low-pass filter. The implemented tracking controller $C_t(s)$ is expressed as follows:
\begin{equation}
    C_t(s) = k_p \cdot \left(1 + \frac{\omega_i}{s}\right)\cdot N_1(s)\cdot N_2(s) \cdot \left(\frac{\omega_{lpf}}{s + \omega_{lpf}}\right),
\end{equation}
where the notch filters $N_j(s)$ for $j\in\{1,2\}$, is as follows: 
\begin{equation}
   \begin{aligned}
       N_j(s) & = \left(\frac{(\frac{s}{\omega_{N_{j}}})^2 + (\frac{s}{Q_{j1}\omega_{N_{j}}}) + 1}{(\frac{s}{\omega_{N_{j}}})^2 + (\frac{s}{Q_{j2}\omega_{N_{j}}}) + 1}\right)
   \end{aligned} 
\end{equation}  

As discussed in Section~\ref{sec:Introduction}, achieving bandwidths beyond those attainable with purely linear control motivates the use of CgLp based on FORE. CgLp provides phase lead at the desired open-loop crossover frequency without altering the loop gain, enabling higher bandwidth while maintaining comparable phase margins. In this work, we employ a proportional GFORE-based CgLp filter with a nonzero feedthrough term ($D_r \neq 0$), which yields an approximately constant gain over the frequency range of interest \cite{hosseini2025frequency}. The CgLp is expressed as: 

\begin{equation}
\operatorname{CgLp}(s)=\mathcal{R}(s)\cdot k_c \cdot C_{l}(s),
\end{equation}
where $\mathcal{R}$ is the proportional FORE with the state-space defined as:
\begin{equation}
\resizebox{\columnwidth}{!}{$
A_r=-\omega_r,\quad B_r=1,\quad C_r=\omega_r,\quad D_r=\frac{\omega_l}{\omega_f-\omega_l},\quad A_\rho=\gamma_r
$}
\end{equation}

and $C_l(s)$ is a lead-lag filter expressed as:
\begin{equation}
C_{l}(s)=\frac{1+\frac{s}{\omega_l}}{1+\frac{s}{\omega_f}}.
\end{equation}
where $\gamma_r \in [-1,1]$ and $\left[\omega_l, \omega_f\right] \in \mathbb{R}_{>0}^{1 \times 2}$ tuned to obtain the desired phase lead within the desired frequency band. The gain correction constant $k_c$ and $\omega_r$ are then calculated as:
\begin{equation}
k_{\mathrm{c}}=\frac{\omega_f-\omega_l}{\omega_f},
\end{equation}
\begin{equation}
\omega_r=\frac{\omega_l}{\sqrt{1+\left(\frac{4\left(1-A_\rho\right)}{\pi\left(1+A_\rho\right)}\right)^2}} .
\end{equation}

In this study, we consider several CgLp configurations that provide phase lead in the range of $5^\circ$–$20^\circ$. To establish a fair benchmark, Case~1 corresponds to a well-tuned purely linear control design. Cases~2–5 augment the tracking controller with CgLp, with phase lead values of $5^\circ$, $10^\circ$, $15^\circ$, and $20^\circ$, respectively. In a later section, Cases~6 and 7 are introduced, which additionally incorporate a shaping filter for the $15^\circ$ and $20^\circ$ phase-lead configurations.

%% file: Text/Simulations.tex
\section{Simulations}
\label{sec:Simulations}

This section compares the baseline linear control design (Case~1) with CgLp-assisted reset control cases that introduce different phase leads at the target crossover frequency to enable higher bandwidth and improved performance. We evaluate the resulting bandwidth and sensitivity improvements and discuss the associated trade-off due to increased higher-order harmonics, which motivates the introduction of a shaping filter. The shaping filter design is then presented, highlighting its effectiveness in reducing error sensitivity and mitigating excessive resetting.

\subsection{Improving Bandwidths using CgLp-based Reset Control}

A baseline linear controller (Case~1) is first designed and tuned to maximize the open-loop crossover frequency $\omega_b$ while ensuring gain and phase margins of at least 6~dB and $60^\circ$, respectively. For CgLp-based cases, the linear damping and tracking controllers are deliberately tuned more aggressively to increase the crossover frequency, resulting in a slightly insufficient phase margin. The CgLp element is then introduced to provide the required phase lead at the target crossover frequency and restore the specified stability margins.

\begin{figure}[t!]
    \centering
    \includegraphics[width=0.85\linewidth]{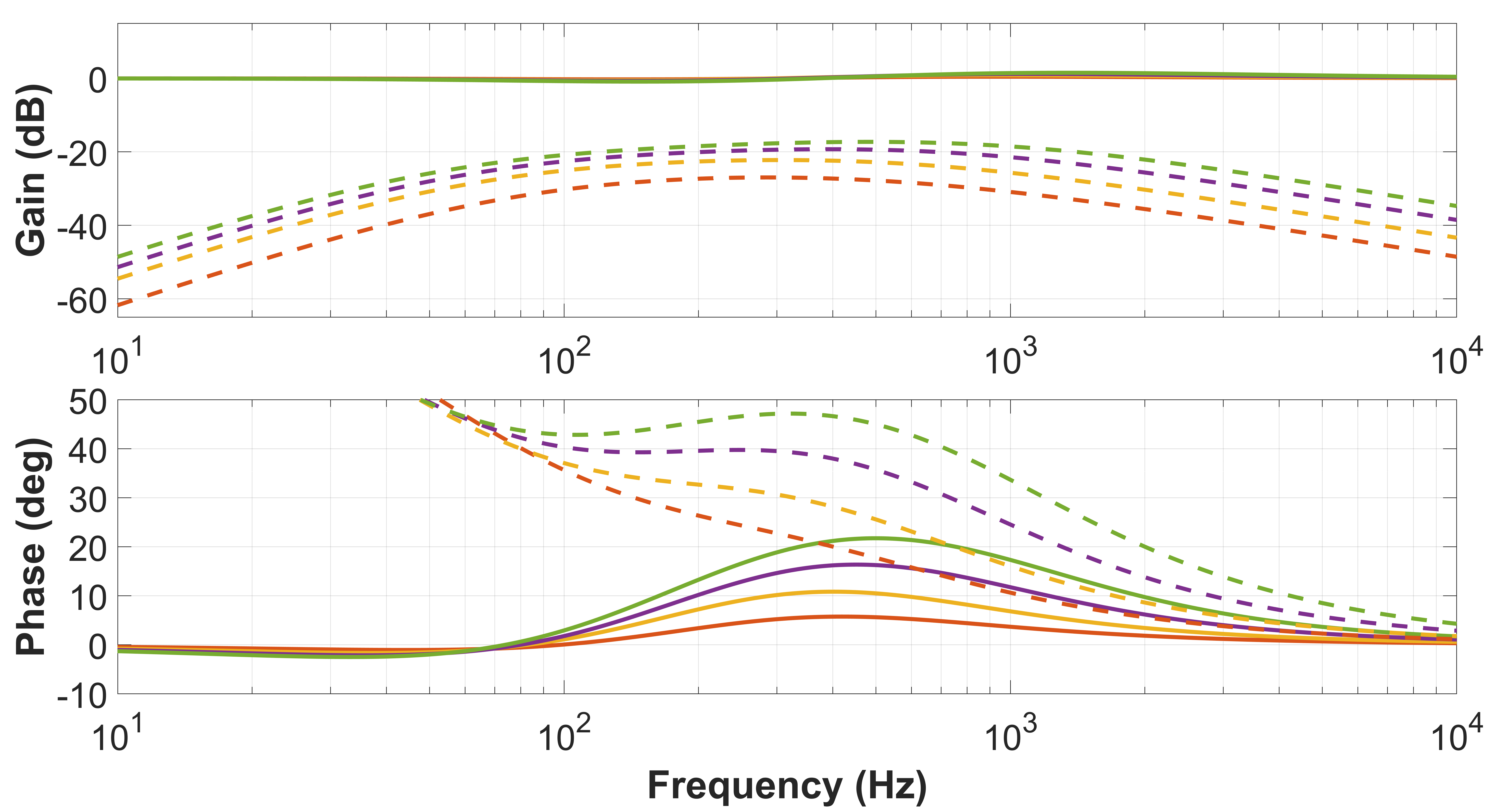}
    \caption{Frequency response from HOSIDF analysis of CgLp with phase lead of $5^\circ$ (\protect\orangeline), $10^\circ$ (\protect\yellowline), $15^\circ$ (\protect\purpleline), and $20^\circ$ (\protect\greenline). Dashed curves indicate the corresponding dominant third-harmonic components.}
    \label{fig:CgLP_HOSIDF}
\end{figure}

Fig.~\ref{fig:CgLP_HOSIDF} shows the HOSIDF of CgLp for phase leads ($\phi_l$) of $5^\circ$, $10^\circ$, $15^\circ$, and $20^\circ$. Around the phase lead frequency, the fundamental component exhibits a small gain dip followed by a rise, with fluctuations of approximately -1 to 2~dB; a larger phase lead increases the gain variation. Below this frequency, the additional phase lag remains small (less than $3^\circ$), and the PI controller and the notch filters are retuned in each case to satisfy the required margins.

For a fair comparison, the NRC damping controller is kept consistent across all cases with $n=8$ and $\gamma=1$, yielding strong suppression of the dominant resonance without amplifying the second mode due to its constant-gain property. The reset jump factor is set to $\gamma_r=0$, providing a practical trade-off between flattening the CgLp first-harmonic response and limiting higher-order harmonic amplification.

Although CgLp can provide a phase lead exceeding the required value, this peak is intentionally not exploited because higher-order harmonics become pronounced near the phase lead region. To limit their influence on low-frequency tracking performance, the peak phase-lead frequency is placed above the targeted crossover frequency in each case (see Fig.~\ref{fig:CgLP_HOSIDF}).

\begin{figure}[t!]
    \centering
    \includegraphics[width=0.85\linewidth]{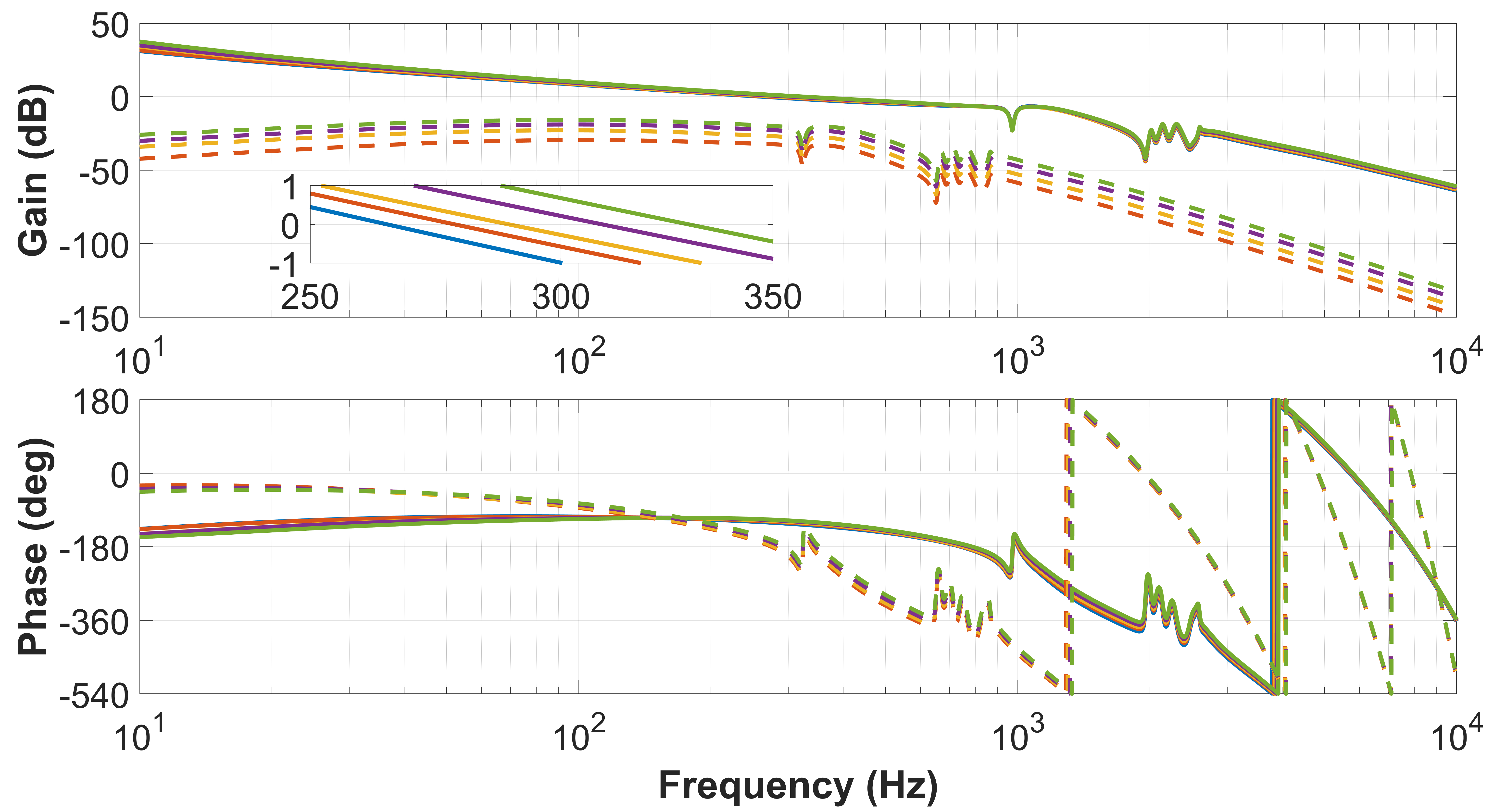}
    \caption{Simulated open-loop frequency response for Case 1 (\protect\blueline), Case 2 (\protect\orangeline), Case 3 (\protect\yellowline), Case 4 (\protect\purpleline), and Case 5 (\protect\greenline). Dashed curves indicate the corresponding dominant third-harmonic components.}
    \label{fig:OL_HOSIDF_Simulation}
\end{figure}

\begin{table}[t]
\caption{CgLp frequencies tuned corresponding to desired phase leads.}
\label{Tab:CgLp_params}
\centering
\begin{tabular}{lcccc}
\hline $\phi_l$ (deg) & ${5}$ & ${10}$ & ${15}$ & ${20}$ \\
\hline$\omega_l~(\mathrm{Hz})$ & 261.6517 & 219.3114 & 205.5480 & 181.2853 \\
$\omega_r~(\mathrm{Hz})$ & 161.6139 & 135.4616 & 126.9604 & 111.9741 \\
$\omega_f~(\mathrm{Hz})$ & 405.6638 & 486.4864 & 672.3213 & 882.7832 \\
\hline
\end{tabular}
\end{table}

\begin{table}[t]
\caption{Simulated open-loop frequency response characteristics.}
\label{Tab:OpenLoop_Sim}
\centering
\resizebox{\columnwidth}{!}{%
\begin{tabular}{lccccccc}
\hline
Case & 1 & 2 & 3 & 4 & 5 & 6 & 7 \\
\hline
$\phi_l$ (deg) & - & $5$ & $10$ & $15$ & $20$ & $15$ & $20$ \\
$\omega_b$ (Hz) & 265.5 & 278.0 & 288.9 & 309.3 & 329.3 & 311.0 & 330.5 \\
PM (deg) & 60.3 & 61.6 & 62.3 & 62.4 & 61.9 & 61.0 & 61.7 \\
GM (dB) & 6.3 & 6.1 & 6.2 & 6.2 & 6.3 & 6.2 & 6.4 \\
\hline
\end{tabular}%
}
\end{table}

Table~\ref{Tab:CgLp_params} lists the tuned CgLp parameters for each case, and Table~\ref{Tab:OpenLoop_Sim} summarizes the simulated open-loop crossover frequencies and the stability margins achieved. In particular, the crossover frequency increases by 65~Hz from the baseline linear design to the highest phase-lead CgLp case while maintaining the required margins (see Fig.~\ref{fig:OL_HOSIDF_Simulation}).

\subsection{Error Sensitivity and Multiple Resetting}

\begin{figure}[t!]
    \centering
    \includegraphics[width=0.85\linewidth]{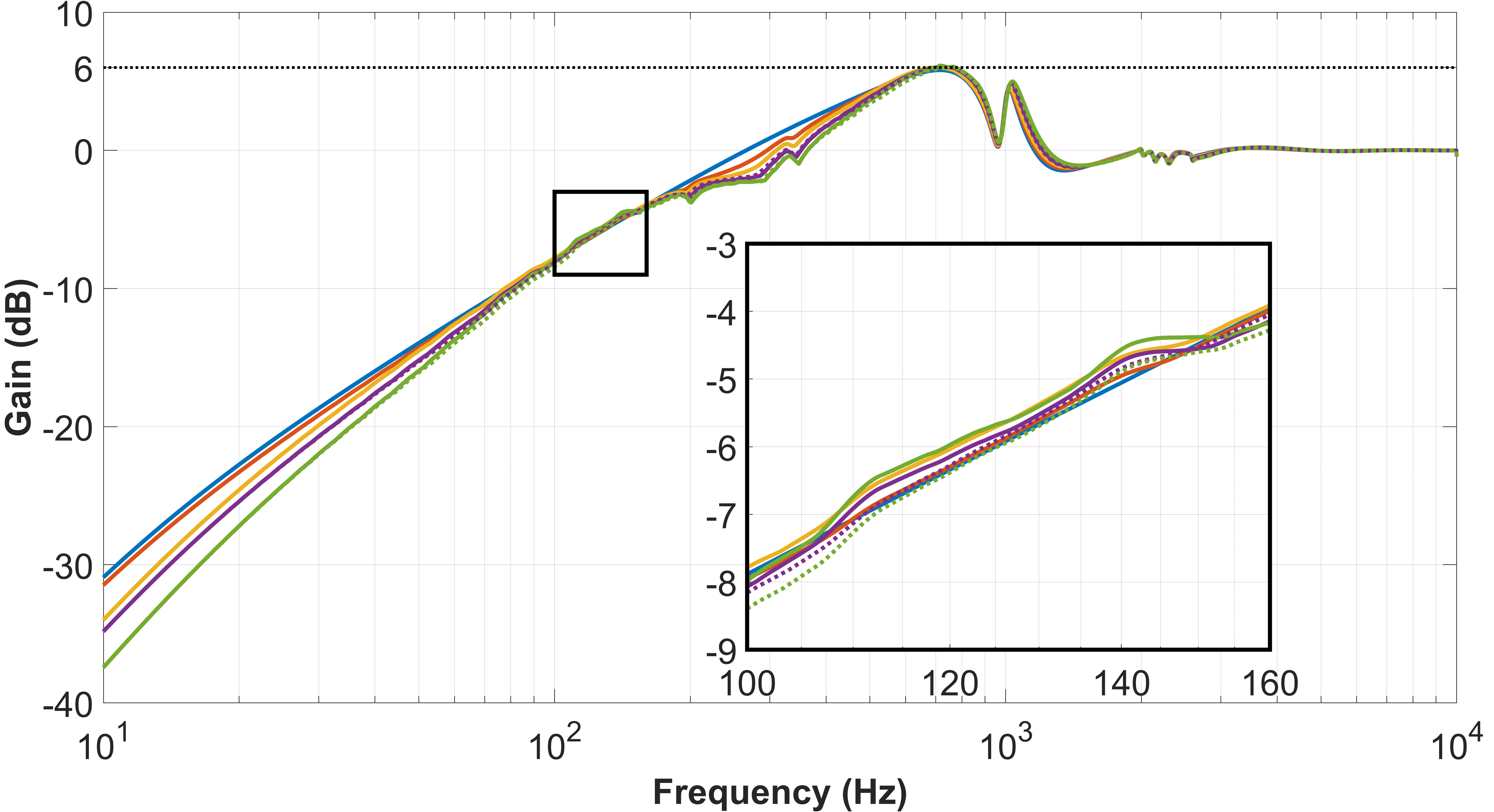}
    \caption{Simulated closed-loop frequency response $S_{er}(s)$ for Case 1 (\protect\orangeline), Case 2 (\protect\orangeline), Case 3 (\protect\yellowline), Case 4 (\protect\purpleline), Case 5 (\protect\greenline), Case 6 (\protect\purplelinedotted), and Case 7 (\protect\greenlinedotted).}
    \label{fig:Sensitvity_Simulation}
\end{figure}

Increasing the CgLp phase lead amplifies higher-order harmonics across the frequency range (Fig.~\ref{fig:CgLP_HOSIDF}). As shown by the error-sensitivity frequency response from the reference $r$ to the control error $e_r$ in Fig.~\ref{fig:Sensitvity_Simulation}, the CgLp-based reset cases achieve lower sensitivity than the baseline linear design over most of the low-frequency band, primarily due to the higher attainable open-loop crossover frequencies. All cases also satisfy the 6~dB sensitivity peak constraint, consistent with the achieved $60^\circ$ phase margins.

\begin{figure}[t!]
    \centering
    \includegraphics[width=0.85\linewidth]{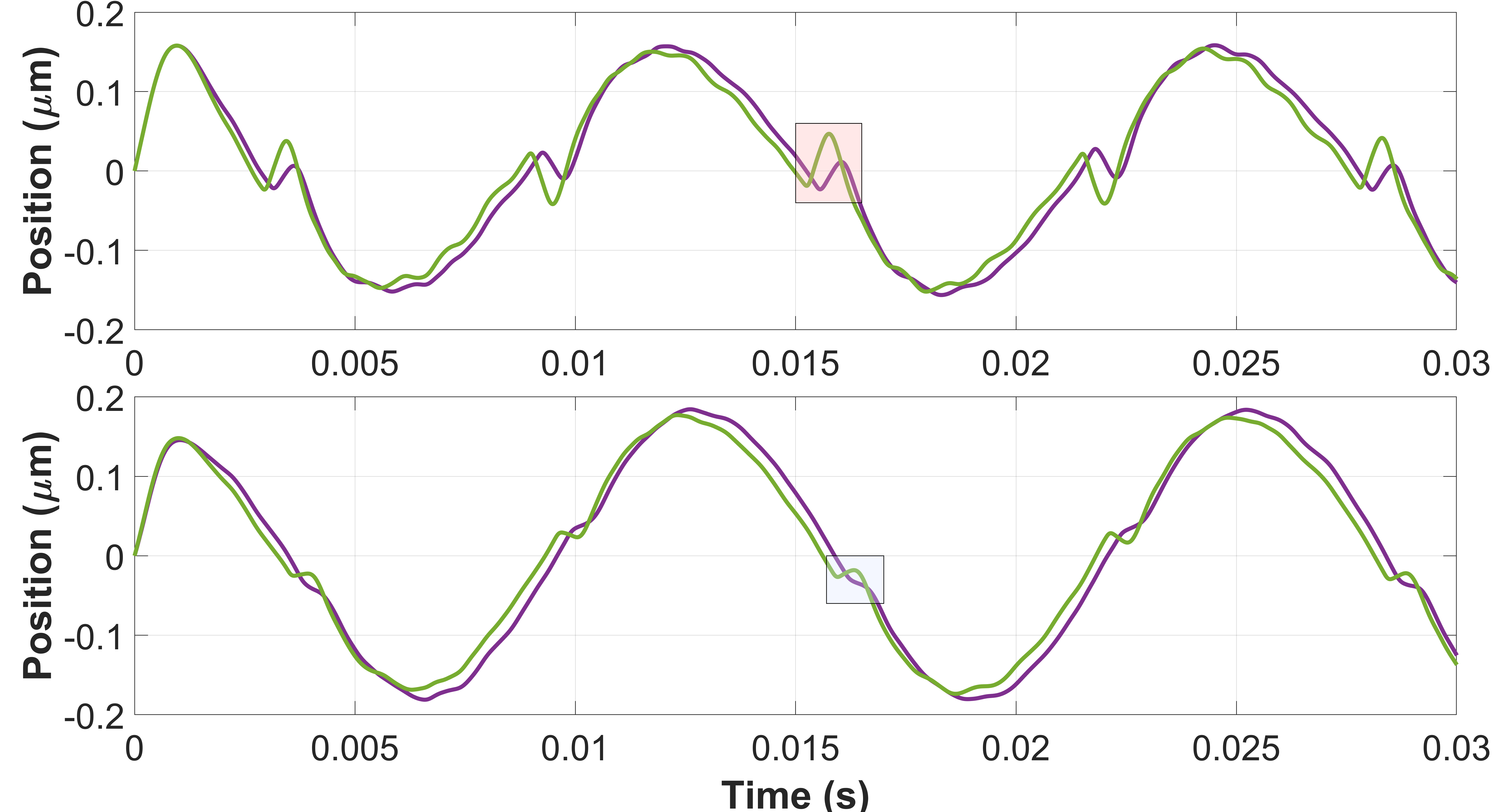}
    \caption{Simulated sinusoidal response at 80~Hz for reset controller: without shaping filter (top plot) - Case 4 (\protect\purpleline) and Case 5 (\protect\greenline), and with shaping filter (bottom plot) - Case 6 (\protect\purpleline) and Case 7 (\protect\greenline).}
    \label{fig:MultipleReset_Simulation}
\end{figure}

Nevertheless, for the higher phase-lead designs, particularly $15^\circ$ and $20^\circ$, the sensitivity exceeds that of the linear baseline in the 80–160~Hz range. This degradation is attributed to the elevated higher-order harmonic content in this band, which becomes comparable to the fundamental response. In the time domain, these harmonics can also induce multiple resets in response to a single-frequency excitation (top plot of Fig.~\ref{fig:MultipleReset_Simulation}). Such behavior violates the HOSIDF assumption of two resets per signal period, rendering the frequency-domain prediction unreliable in this range. To address these issues, a shaping filter is introduced in the next subsection.

\subsection{Attenuating Nonlinearities via Shaping Filter}

To mitigate the higher-order harmonics in the problematic frequency range, a shaping filter is introduced to regulate the reset triggering signal $e_s$ applied to the reset element. Following \cite{zhang2025enhancing}, the shaping filter is placed in the reset path to impose a controlled phase difference between the error signal $e$ and the triggering signal $e_s$. Since this modification affects only the reset action, it can attenuate higher-order harmonics without directly altering the linear loop dynamics.

In this work, the shaping filter $C_s(s)$ is defined as
\begin{equation}
C_{s}(s)=\frac{\mathcal{R}_{\mathrm{bl}}(s)}{N_{s_1}(s)\cdot N_{s_2}(s)\cdot C_L(s)},
\end{equation}
where $\mathcal{R}_{\mathrm{bl}}(s)$ is the base linear model of the PFORE element. The notch filters $N_{s_1}(s)$ and $N_{s_2}(s)$, and the fractional lead--lag filter $C_L(s)$ are
\begin{equation}
\resizebox{\columnwidth}{!}{$
\begin{aligned}
N_{s_1}(s) &=
\frac{\left(\frac{s}{\omega_L}\right)^2+\left(\frac{s}{\omega_L}\right)+1}
{\left(\frac{s}{\omega_L}\right)^2+\left(\frac{s}{Q\omega_L}\right)+1},\quad
N_{s_2}(s) &=
\frac{\left(\frac{s}{\omega_H}\right)^2+\left(\frac{s}{Q\omega_H}\right)+1}
{\left(\frac{s}{\omega_H}\right)^2+\left(\frac{s}{\omega_H}\right)+1}
\end{aligned}
$}
\end{equation}

\begin{equation}
C_L(s)=\left(\frac{1+s/\omega_L}{1+s/\omega_H}\right)^\lambda .
\end{equation}
The notch filters are positioned at the bounds of the intended phase-lead interval $[\omega_L,\omega_H]$, while the fractional exponent $\lambda$ provides additional tuning to shape the phase response within this fixed frequency range.

For implementation, the fractional-order term is approximated by a rational transfer function using MATLAB’s \texttt{invfreqs} (nonlinear least-squares fitting). A second-order approximation is selected to balance approximation fidelity against numerical robustness in implementation.

\begin{figure}[t!]
    \centering
    \includegraphics[width=0.85\linewidth]{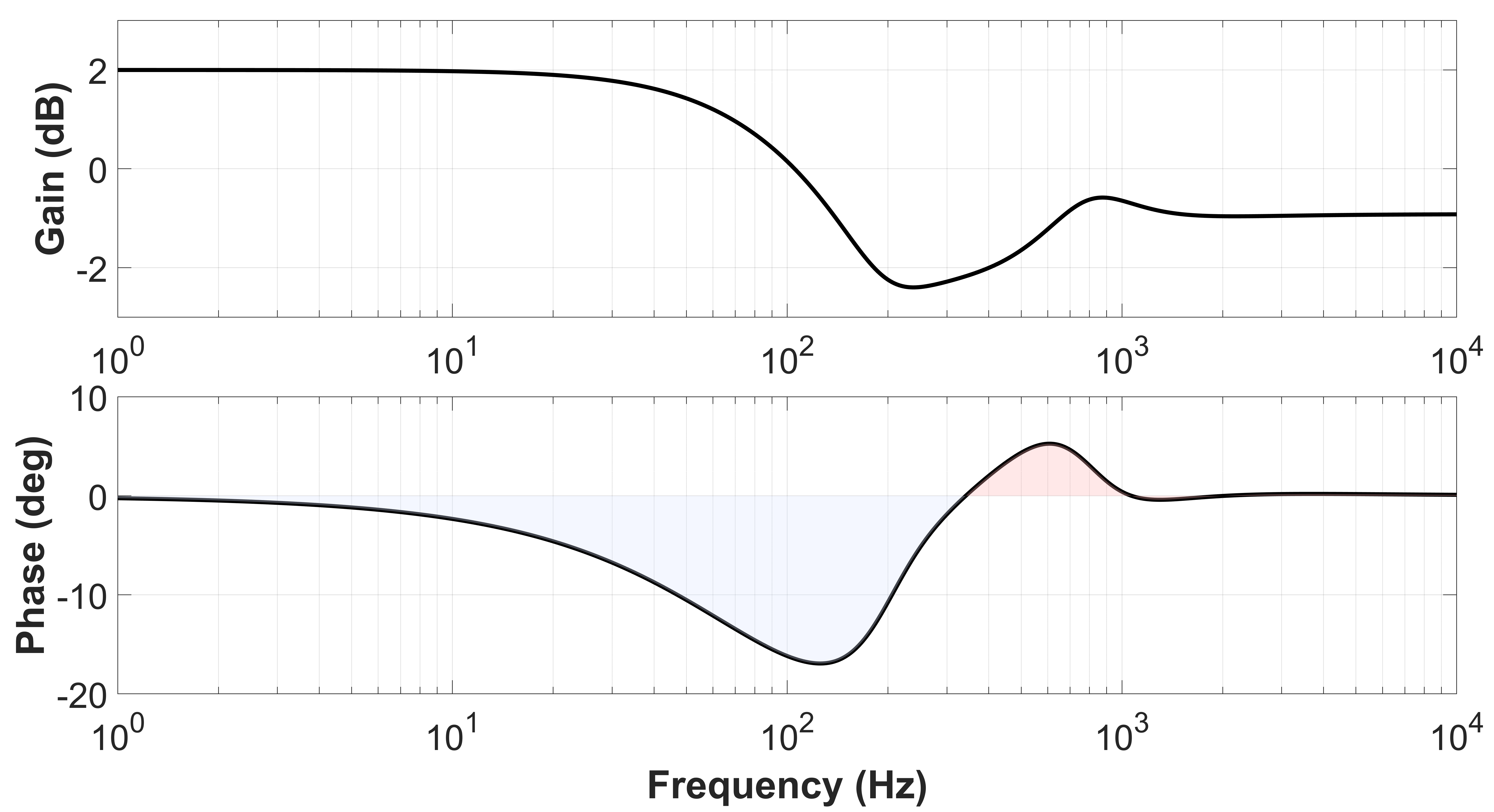}
    \caption{Frequency response of tuned shaping filter $C_s(s)$ (\protect\blackline).}
    \label{fig:ShapingFilter}
\end{figure}

\begin{figure}[t!]
    \centering
    \includegraphics[width=0.85\linewidth]{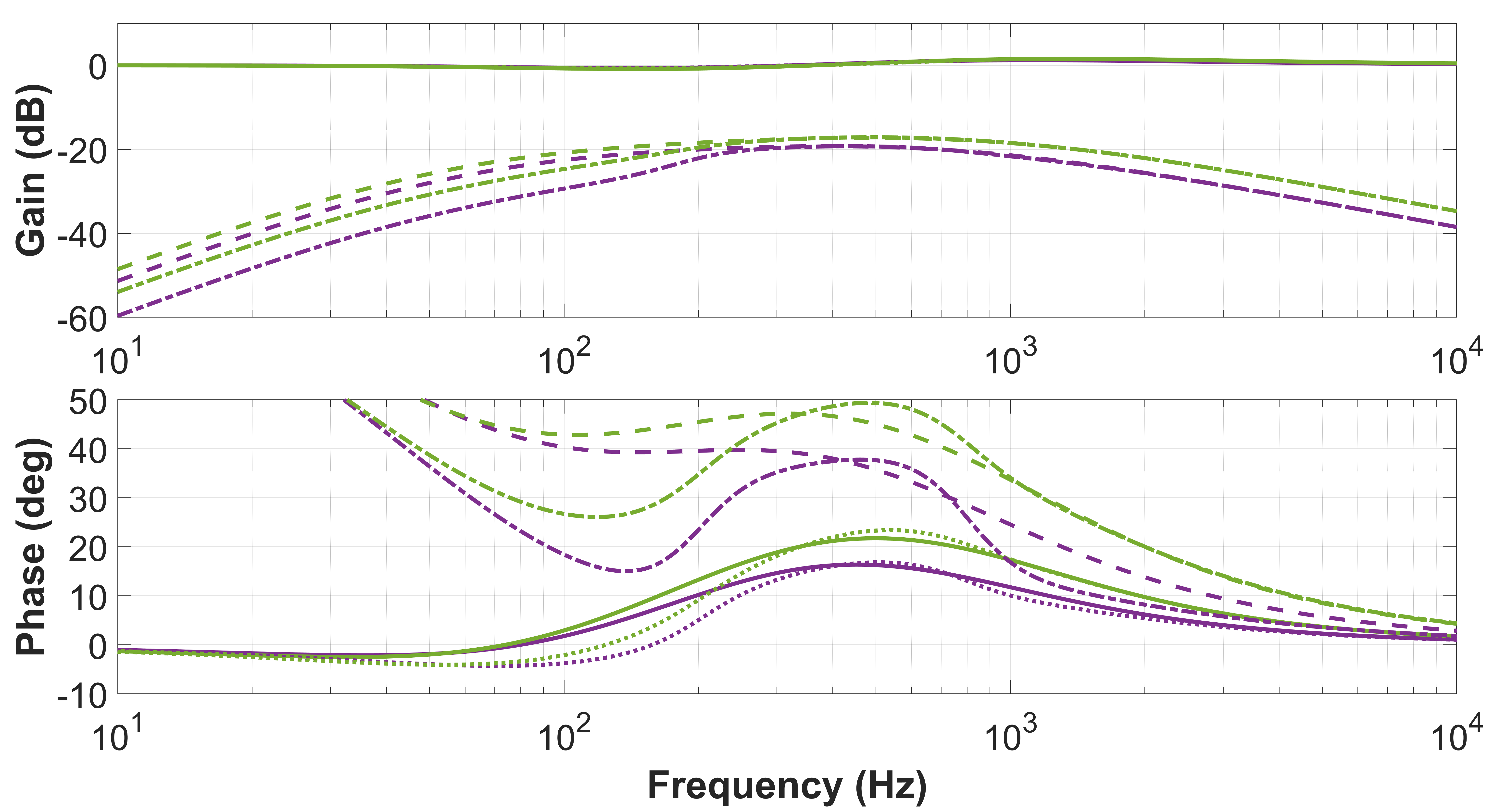}
    \caption{HOSIDF-based frequency response of higher-order harmonics: without shaping filter - Case 4 (\protect\purplelinedashed) and Case 5 (\protect\greenlinedashed), and with shaping filter - Case 6 (\protect\purplelinedashdot) and Case 7 (\protect\greenlinedashdot).}
    \label{fig:CgLp_HOSIDF_SF}
\end{figure}

\begin{table}[t!]
\centering
\caption{Shaping filter parameters tuned for Cases 6 and 7.}
\label{tab:ShapingFilter}
\begin{tabular}{cccccc}
\hline Case & $\phi_l$ & $\omega_L(\mathrm{~Hz})$ & $\omega_H(\mathrm{~Hz})$ & $\lambda$ & $Q$ \\
\hline 6 & $15^\circ$ & 200 & 800 & -0.4 & 1.3 \\
7 & $20^\circ$ & 200 & 800 & -0.9 & 1.15 \\
\hline
\end{tabular}
\end{table}

The shaping-filter parameters (Table~\ref{tab:ShapingFilter}) are tuned for the $15^\circ$ (Case~6) and $20^\circ$ (Case~7) CgLp designs to introduce a phase lag below 200~Hz (light-blue region in Fig.~\ref{fig:ShapingFilter}). This additional lag reduces low-frequency higher-order harmonics while preserving the required phase lead at the target crossover frequency and avoiding unintended changes at higher frequencies. As shown in Fig.~\ref{fig:CgLp_HOSIDF_SF}, the dominant third-harmonic component is substantially reduced below 200~Hz.

This phase shaping also improves the error sensitivity: in Fig.~\ref{fig:Sensitvity_Simulation}, the elevated sensitivity observed in the 80–160~Hz band for the higher phase-lead cases is reduced to approximately the baseline linear level. In the time domain, the sinusoidal response in the bottom plot of Fig.~\ref{fig:MultipleReset_Simulation} confirms that the shaping filter suppresses the higher-order harmonic-induced multiple-reset behavior, restoring the intended two resets per cycle without degrading the fundamental tracking response.

%% file: Text/Experimental.tex
\section{Experimental Results}
\label{sec:Experimental}

This section presents the experimental implementation of the simulated cases, including the shaping-filter designs, and evaluates their performance using frequency-domain sensitivity measurements and time-domain tracking results.

\subsection{Experimental Frequency Responses}

Due to small discrepancies between the identified and estimated frequency responses, the damping and tracking controllers are slightly retuned from the simulated cases to achieve the desired open- and closed-loop behavior. The implemented parameters are summarized in Table~\ref{tab:controller_params}. The controllers are then discretized using the bilinear transform and implemented on the FPGA in closed loop.

\begin{table}[t]
\centering
\caption{Tracking controller parameters for experimental implementation.}
\label{tab:controller_params}
\renewcommand{\arraystretch}{1.1}
\resizebox{\columnwidth}{!}{%
\begin{tabular}{lccccccc}
\hline
\textbf{Parameters} & \textbf{Case 1} & \textbf{Case 2} & \textbf{Case 3} & \textbf{Case 4} & \textbf{Case 5} & \textbf{Case 6} & \textbf{Case 7} \\
\hline
$k_p$ & 0.9879 & 1.0537 & 1.1028 & 1.2129 & 1.2129 & 1.2129 & 1.2129 \\
\hline
$\omega_i\,(\mathrm{Hz})$ & 10 & 10 & 15 & 15 & 15 & 15 & 15 \\
\hline
$\omega_{N_1}\,(\mathrm{Hz})$ & 1100.0 & 1100.0 & 1050.0 & 1000.0 & 1000.0 & 1000.0 & 1000.0 \\
\hline
$Q_{11}$ & 1.0500 & 0.9680 & 1.0125 & 1.0571 & 1.0571 & 1.0571 & 1.0571 \\
\hline
$Q_{12}$ & 1.0000 & 0.8000 & 0.7500 & 0.7000 & 0.7000 & 0.7000 & 0.7000 \\
\hline
$\omega_{N_2}\,(\mathrm{Hz})$ & 2582.0 & 2582.0 & 2582.0 & 2582.0 & 2582.0 & 2582.0 & 2582.0 \\
\hline
$Q_{21}$ & 40.0 & 40.0 & 40.0 & 40.0 & 40.0 & 40.0 & 40.0 \\
\hline
$Q_{22}$ & 5.0 & 5.0 & 5.0 & 5.0 & 5.0 & 5.0 & 5.0 \\
\hline
$\omega_{{lpf}}\,(\mathrm{Hz})$ & 5000.0 & 5000.0 & 5000.0 & 5000.0 & 5000.0 & 5000.0 & 5000.0 \\
\hline
\end{tabular}%
}
\end{table}

\begin{figure}[t!]
    \centering
    \includegraphics[width=0.85\linewidth]{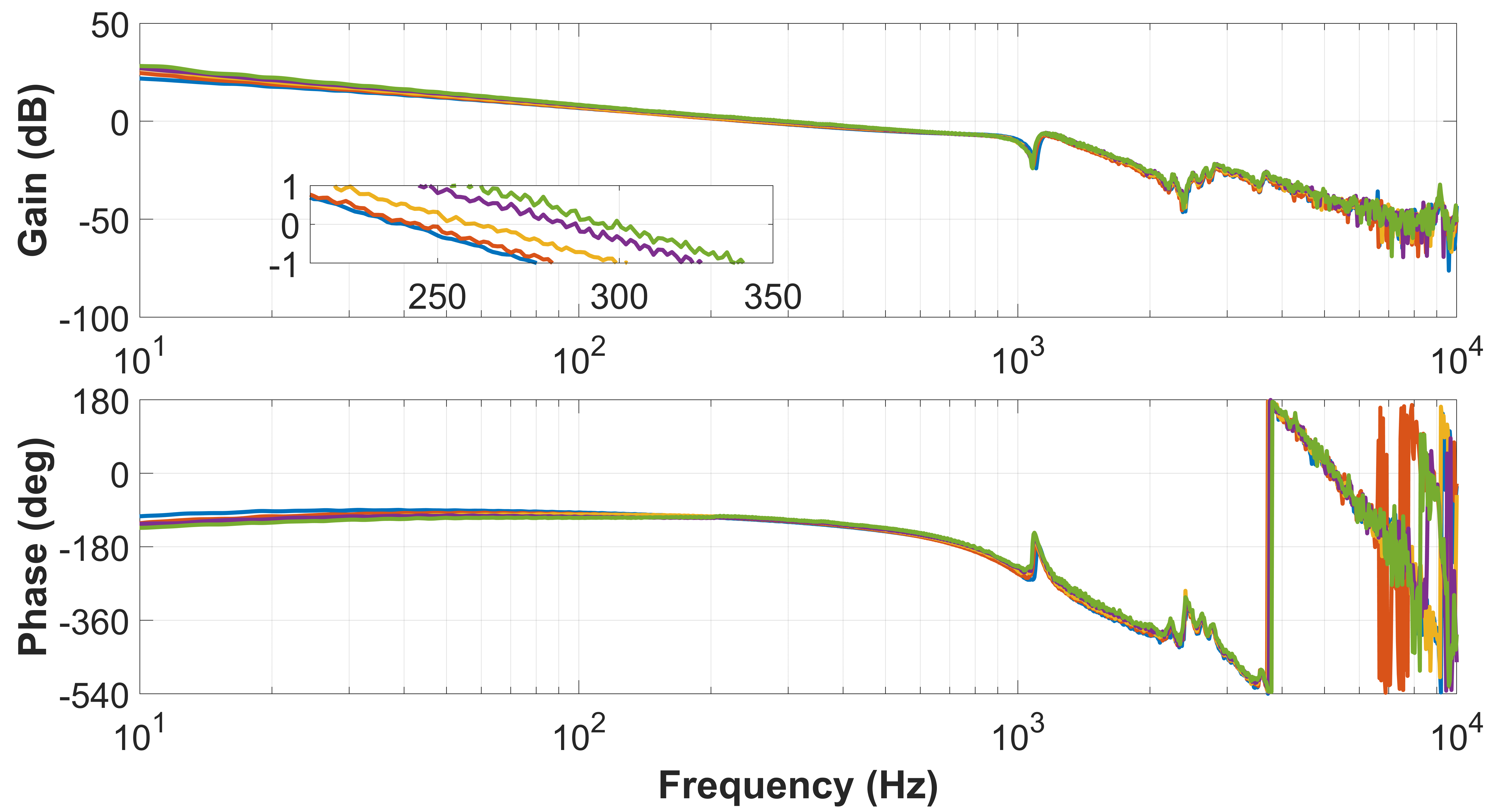}
    \caption{Experimental open-loop frequency response for Case 1 (\protect\blueline), Case 2 (\protect\orangeline), Case 3 (\protect\yellowline), Case 4 (\protect\purpleline), and Case 5 (\protect\greenline).}
    \label{fig:OpenLoop_Experimental}
\end{figure}

\begin{figure}[t!]
    \centering
    \includegraphics[width=0.85\linewidth]{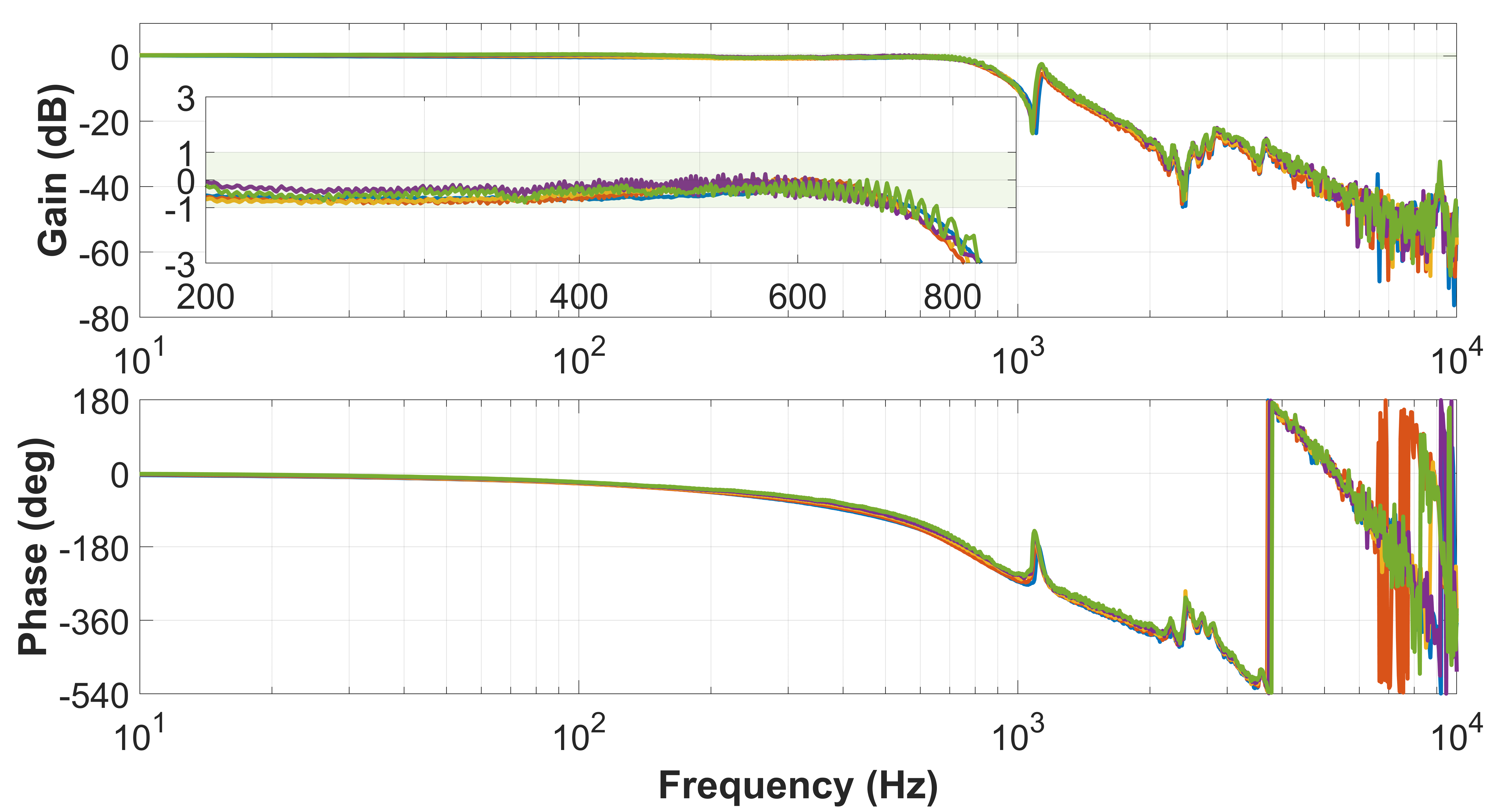}
    \caption{Experimental closed-loop frequency response $T_{yr}(s)$ for Case 1 (\protect\blueline), Case 2 (\protect\orangeline), Case 3 (\protect\yellowline), Case 4 (\protect\purpleline), and Case 5 (\protect\greenline).}
    \label{fig:ClosedLoop_Experimental}
\end{figure}

\begin{figure}[t!]
    \centering
    \includegraphics[width=0.85\linewidth]{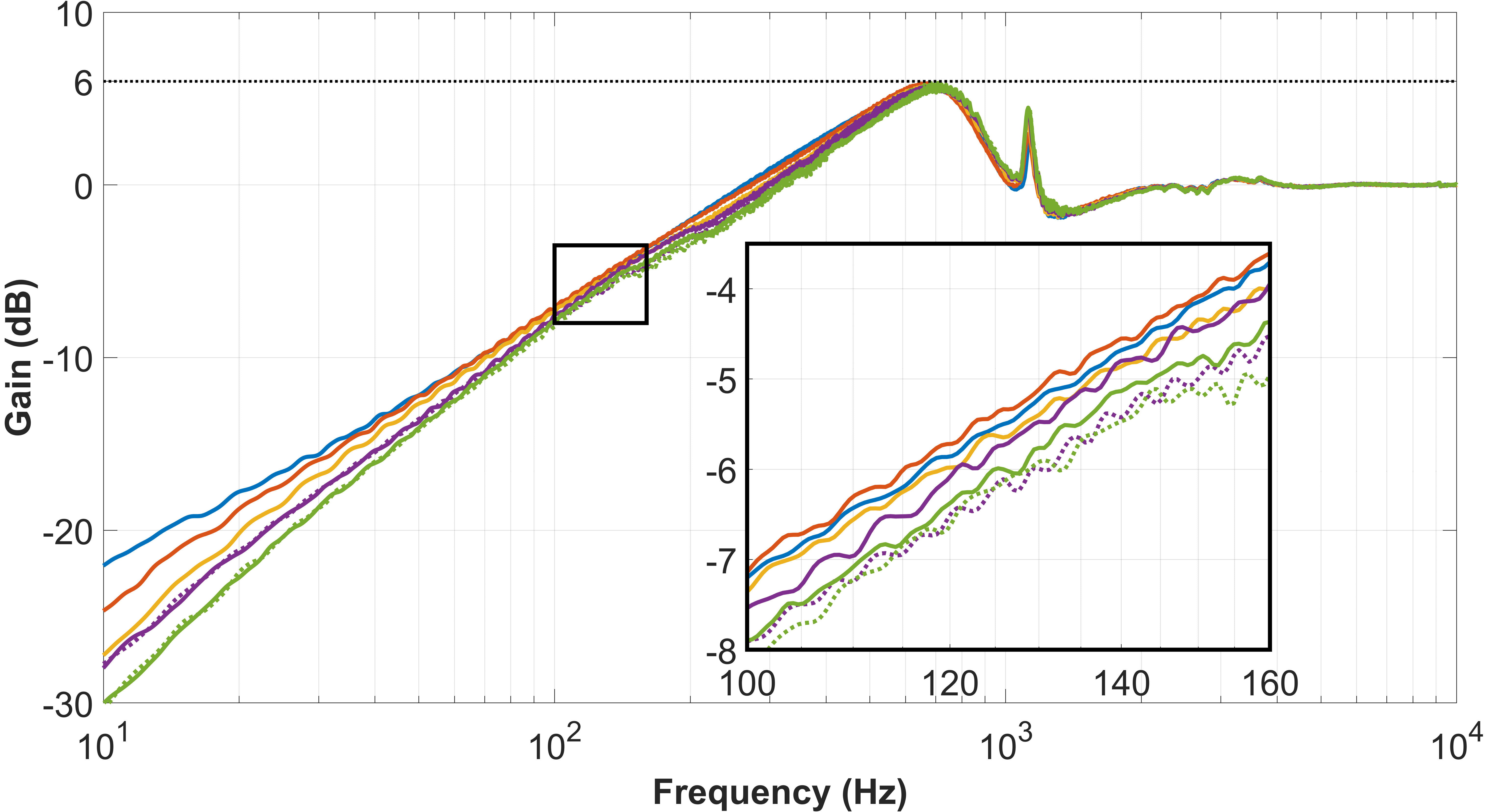}
    \caption{Experimental closed-loop frequency response $T_{yr}(s)$ for Case 1 (\protect\blueline), Case 2 (\protect\orangeline), Case 3 (\protect\yellowline), Case 4 (\protect\purpleline), Case 5 (\protect\greenline), Case 6 (\protect\purplelinedotted), and Case 7 (\protect\greenlinedotted).}
    \label{fig:Sensitivity_Experimental}
\end{figure}

\begin{table}[t]
\centering
\caption{Experimental open and closed-loop frequency response characteristics.}
\label{tab:freq_response_chars}
\renewcommand{\arraystretch}{1.1}
\resizebox{\columnwidth}{!}{%
\begin{tabular}{lccccccc}
\hline
Case & 1 & 2 & 3 & 4 & 5 & 6 & 7 \\
\hline
$\phi_l$ (deg) & - & $5$ & $10$ & $15$ & $20$ & $15$ & $20$ \\
$\omega_b$ (Hz) & 270.4 & 274.2 & 292.4 & 310.5 & 325.1 & 313.3 & 325.1 \\
PM (deg) & 60.6 & 62.1 & 61.2 & 61.5 & 62.2 & 60.0 & 61.7 \\
GM (dB) & 6.3 & 6.3 & 6.6 & 6.7 & 6.6 & 6.7 & 6.6 \\
$\omega_c$ (Hz) & 809.1 & 805.4 & 824.1 & 827.9 & 843.3 & 827.9 & 843.3 \\
\hline
\end{tabular}%
}
\end{table}

Following the tuning procedure in Section~\ref{sec:Simulations}, the tracking controllers are designed to increase the open-loop crossover frequency relative to the baseline linear case, with the resulting phase-margin shortfall compensated by CgLp filters providing different phase leads. The measured open-loop frequency responses in Fig.~\ref{fig:OpenLoop_Experimental} confirm the intended crossover-frequency increase. The results in Table~\ref{tab:freq_response_chars} indicate an open-loop bandwidth improvement of approximately 55~Hz.

Figure~\ref{fig:ClosedLoop_Experimental} shows the measured closed-loop frequency response mapping from the reference $r$ to the measured output $y$. The closed-loop bandwidth $\omega_c$, defined by the $\pm3$~dB crossings, improves by about 34~Hz. With the NRC damping controller, bandwidths can be achieved beyond the first dominant resonance, and the reset-based designs extend the bandwidth further.

Finally, the error sensitivity is measured using the base linear approximation and is shown in Fig.~\ref{fig:Sensitivity_Experimental}. The increased bandwidth allows a higher integrator corner frequency in the tracking controller, reducing sensitivity over most of the frequency range of interest. However, sensitivity increases in the 80–160~Hz band due to higher-order harmonic effects. Implementing the shaping filter mitigates this degradation by attenuating the influence of these harmonics.

\subsection{Tracking Performance}

\begin{figure*}[t!]
	\centering
	\subfloat[]{\includegraphics[width = 0.42\textwidth]{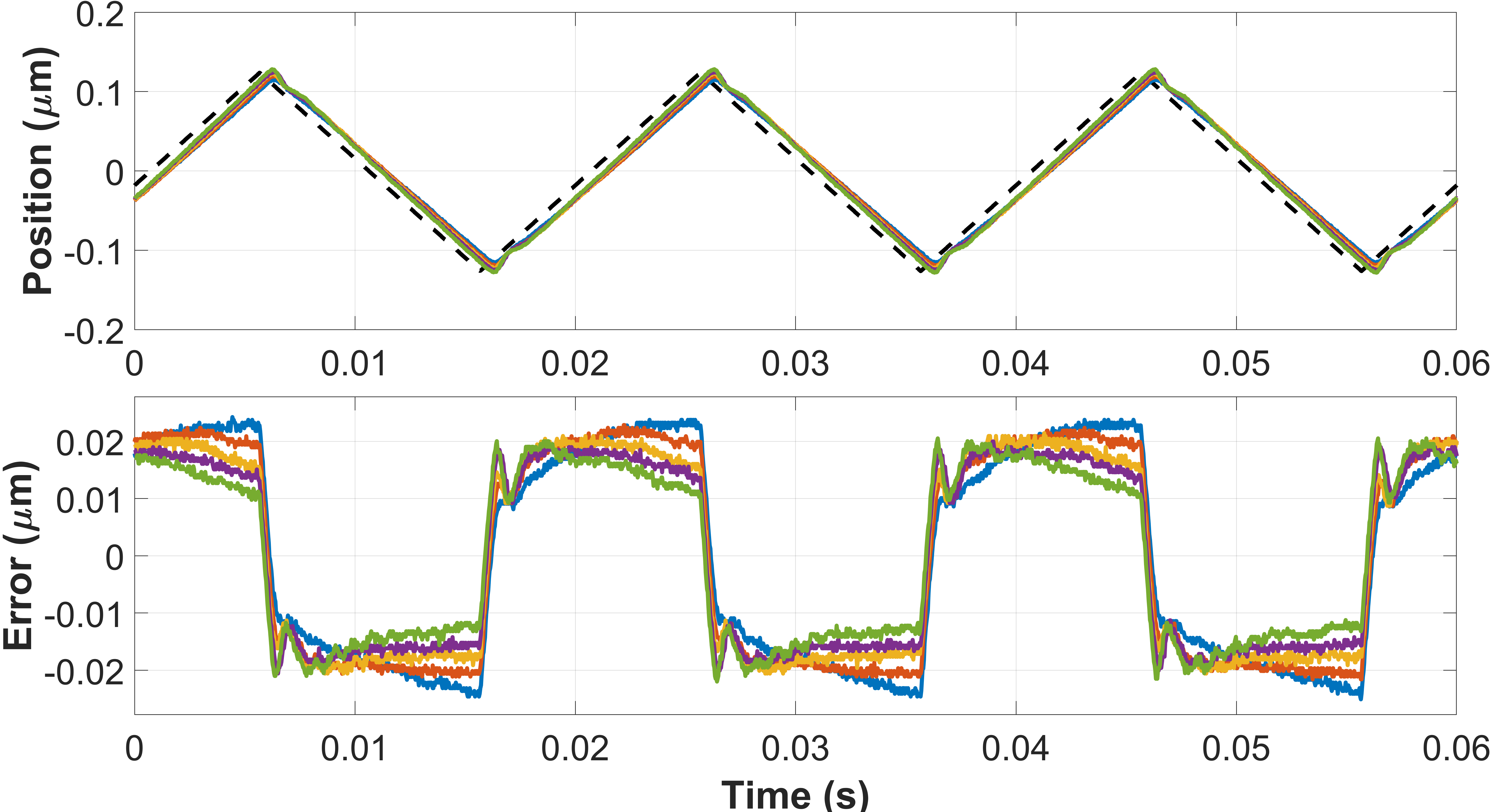}\label{fig:triangular50}}
	 \hspace{1mm}
	\subfloat[]{\includegraphics[width = 0.42\textwidth]{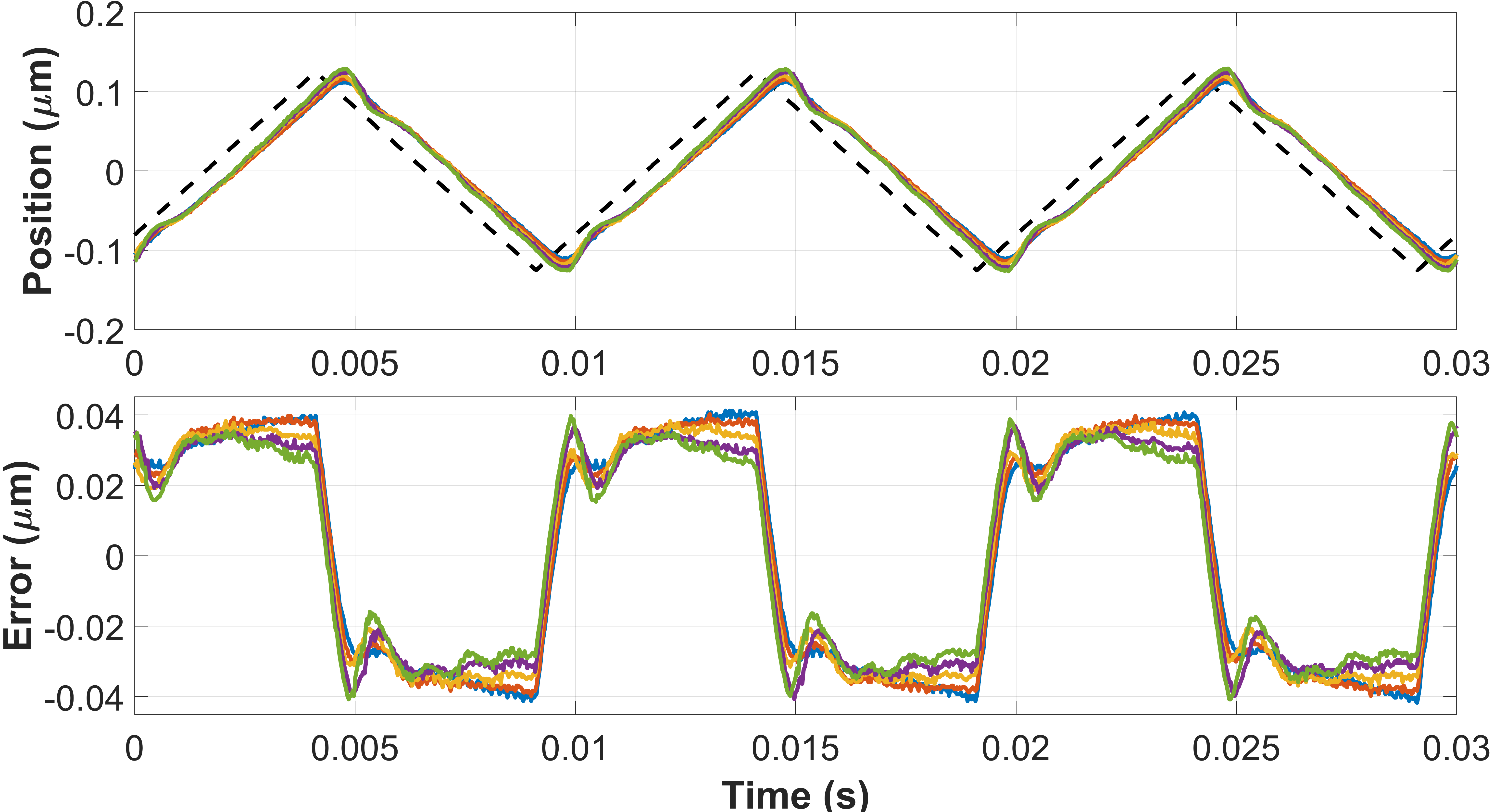}\label{fig:triangular100}}
    \hfill
        \caption{Experimental triangular reference tracking output (top plot) and error (bottom plot) without phase delay compensation for Case 1 (\protect\blueline), Case 2 (\protect\orangeline), Case 3 (\protect\yellowline), Case 4 (\protect\purpleline), and Case 5 (\protect\greenline): (a) 50 Hz, and (b) 100 Hz.}
        \label{fig:TriangularTracking}
\end{figure*}

To evaluate tracking performance, triangular reference trajectories, commonly used in raster-scanning applications, are applied with fundamental frequencies from 1 to 300~Hz. Figure~\ref{fig:TriangularTracking} shows representative tracking responses and the corresponding errors (without delay compensation) at selected frequencies. For brevity, the shaping-filter cases are omitted, as their time-domain differences are visually negligible.

The tracking accuracy is quantified using the root-mean-square (RMS) error across the tested reference frequencies (Table~\ref{tab:rms_error}). The results show a consistent reduction in RMS error for CgLp-assisted reset cases compared to the baseline linear controller, with improved performance as the designed phase lead increases. This trend is consistent with the measured error-sensitivity responses. For higher phase-lead designs, the shaping filter further reduces higher-order harmonic effects and yields an additional reduction in RMS error compared to the corresponding cases without shaping.

\begin{table}[t!]
\centering
\caption{RMS tracking error ($\mu$m) for different triangular reference frequencies.}
\label{tab:rms_error}
\renewcommand{\arraystretch}{1.1}
\resizebox{\columnwidth}{!}{%
\begin{tabular}{c|ccccccc}
\hline
\textbf{Frequency (Hz)} & \textbf{Case 1} & \textbf{Case 2} & \textbf{Case 3} & \textbf{Case 4} & \textbf{Case 5} & \textbf{Case 6} & \textbf{Case 7} \\
\hline
1   & 0.0010 & 0.0007 & 0.0007 & 0.0007 & 0.0006 & 0.0007 & 0.0006 \\
5   & 0.0042 & 0.0027 & 0.0020 & 0.0019 & 0.0018 & 0.0016 & 0.0014 \\
10  & 0.0066 & 0.0053 & 0.0041 & 0.0037 & 0.0037 & 0.0030 & 0.0030 \\
20  & 0.0099 & 0.0093 & 0.0080 & 0.0072 & 0.0073 & 0.0063 & 0.0064 \\
50  & 0.0186 & 0.0186 & 0.0175 & 0.0158 & 0.0162 & 0.0150 & 0.0152 \\
100 & 0.0322 & 0.0326 & 0.0311 & 0.0288 & 0.0297 & 0.0278 & 0.0287 \\
150 & 0.0453 & 0.0450 & 0.0436 & 0.0412 & 0.0426 & 0.0397 & 0.0411 \\
200 & 0.0570 & 0.0568 & 0.0543 & 0.0512 & 0.0526 & 0.0495 & 0.0508 \\
300 & 0.0789 & 0.0761 & 0.0725 & 0.0701 & 0.0713 & 0.0666 & 0.0676 \\
\hline
\end{tabular}%
}
\end{table}

%% file: Text/Conclusions.tex
\section{Conclusion}
\label{sec:Conclusion}

This paper addressed the bandwidth and tracking limitations of piezoelectric nanopositioning systems by integrating nonlinear reset control with resonance-targeted active damping. A dual-loop controller was developed in which an NRC-based inner loop suppresses dominant resonant dynamics, enabling an outer-loop tracking controller with a CgLp reset element to recover phase at the targeted crossover and achieve higher bandwidth without high-frequency gain amplification. We identified that higher phase-lead CgLp settings can amplify higher-order harmonics, leading to localized sensitivity degradation and multiple resetting, and we mitigated this effect using a shaping filter placed in the reset-trigger path to regulate reset timing and attenuate harmonic influence. Real-time experiments on an industrial nanopositioner confirmed the effectiveness of the approach, yielding approximately 55~Hz improvement in open-loop crossover frequency and about 34~Hz increase in closed-loop bandwidth compared with a baseline linear design, while maintaining the intended robustness margins.
\section*{}

%% file: Text/Statements.tex
\section*{Acknowledgments}
This work was financed by Physik Instrumente (PI) SE \& Co. KG and co-financed by Holland High Tech with PPS Project supplement for research and development in the field of High Tech Systems and Materials.
